\documentclass[aps,pro,superscriptaddress,twocolumn,floatfix]{revtex4}
\usepackage{amsfonts,color}
\usepackage{amsmath}
\usepackage{amssymb}
\usepackage{graphicx,dsfont}

\def\im{{\sf {i}}}

\begin{document}

\title{Many-body characterization of topological superconductivity: \\ The Richardson-Gaudin-Kitaev chain}
\author{Gerardo Ortiz}
\affiliation{Department of Physics, Indiana University, Bloomington,
IN 47405, USA}
\email{ortizg@indiana.edu}
\author{Jorge Dukelsky}
\affiliation{Instituto de Estructura de la Materia, C.S.I.C., Serrano 123,
E-28006 Madrid, Spain}
\author{Emilio Cobanera}
\affiliation{Instituut-Lorentz, Universiteit Leiden, P.O. Box 9506, 2300 RA Leiden,
The Netherlands}
\author{Carlos Esebbag}
\affiliation{Departamento de F\'{i}sica y Matem\'{a}ticas, Universidad de
Alcala, 28871 Alcala de Henares, Spain}
\author{Carlo Beenakker}
\affiliation{Instituut-Lorentz, Universiteit Leiden, P.O. Box 9506, 2300 RA Leiden,
The Netherlands}
\begin{abstract}
What distinguishes trivial from topological superfluids in interacting many-body systems
where the number of particles is conserved? Building on a class of integrable pairing Hamiltonians, we 
present a number-conserving, interacting variation of the Kitaev model, the Richardson-Gaudin-Kitaev chain, 
that remains exactly solvable for periodic and antiperiodic boundary conditions. Our 
model allows us to identify fermion parity switches that distinctively 
characterize topological superconductivity in interacting many-body systems. Although the Majorana 
zero-modes in this model have only a power-law confinement, we may still define many-body 
Majorana operators by tuning the flux to a fermion parity switch. 
We derive a closed-form expression for an interacting topological invariant and show that the transition 
away from the topological phase is of third order.
\end{abstract}
\date{July 2014}

\maketitle

In recent years, the physics of Majorana zero-energy modes has become a
key subfield of condensed matter physics \cite{Has10,Qi11,Alicea2012,Beenakker2013}. 
On the theory side, a central result is the bulk-boundary correspondence 
\cite{Ryu2010} that associates Majorana zero-modes to the boundary of
(or defects in) a topologically non-trivial superconductor, with the Kitaev chain as a prototypical example \cite{Kitaev2001}.
The mathematical formalism underlying this correspondence relies on the symmetries and topological
invariants of the Bogoliubov-de Gennes equation \cite{DeG66}, a mean-field description of the superconducting state in which the 
conservation of the number of fermions (a continuous symmetry) is broken down to a 
discrete symmetry, the conservation of fermion-number parity. Majorana zero-modes 
are directly linked to the spontaneous breaking of this residual discrete
symmetry \cite{vanHeck2014}. 

As the experimental side of Majorana physics continues to develop
\cite{Mourik2012,Das2012,Deng2012,Fin13,Chu13,Sta13}, it becomes crucial to unveil how much of
the mean-field picture survives beyond its natural limits.
This has motivated recent studies 
\cite{Tsvelik,Cheng2011,Hassler2012,Manolescu2014,Sau2011,Fidkowski2011,vanHeck2011}, 
with the focus on the anomalous $2\Phi_0=h/e$ flux-periodicity of the 
Josephson effect --- the hallmark of a topological superconductor 
\cite{Kitaev2001}.

A main thrust of this paper is the characterization of  interacting many-body, number-conserving, topological
superconductors, or superfluids, and the fate of Majorana zero-modes beyond mean-field. 
The theoretical study of any interacting quantum system
is hampered by the exponential growth of the Hilbert space with the number of particles. 
An additional complication of superconducting systems is the lack of simple principles to guide the
design of particle-number conserving models, in which 
the phase of the order parameter is not a good quantum number. 
To overcome both obstacles, we have constructed an
exactly solvable, number-conserving variation of the Kitaev
chain. Because our model belongs to a class of integrable pairing models \cite{Amico2001,Dukelsky2001,ortiz2005} 
based on the $s$-wave reduced BCS Hamiltonian first solved by Richardson \cite{Richardson1963},  and on a generalization 
of the Gaudin magnet \cite{Gaudin1976}, 
we will refer to it as the Richardson-Gaudin-Kitaev (RGK) chain.

The RGK model is integrable for periodic and antiperiodic boundary conditions. It
allows us to obtain precise answers for the characterization problem posed here, including  the very existence 
of a topologically nontrivial phase, in an interacting number-conserving system, the order of the 
phase transition into the trivial phase, the definition of a topological invariant beyond mean-field, 
and the emergence of many-body Majorana zero-modes.

{\it Richardson-Gaudin-Kitaev chain.---} 
Our model is defined by the Hamiltonian 
\begin{eqnarray}
H_{\sf RGK}=\!\!\!\sum_{k\in  {\cal S}_{k}^{\phi}}
\varepsilon_k \, \hat{c}^\dagger_k\hat{c}_k^{\;}
- 8 G\hspace*{-0.3cm}\sum_{k,k' \in {\cal S}_{k+}^{\phi}}\hspace*{-0.25cm}
\eta_k \eta_{k'}
\hat{c}_{k}^\dagger \hat{c}_{-k}^\dagger \hat{c}_{-k'}^{\;}\hat{c}_{k'}^{\;}
\label{HamiltRGnew}
\end{eqnarray}
in terms of fermion creation operators $ \hat{c}^\dagger_k$, with momentum $k$-dependent single-particle
spectrum $\varepsilon_k=-2t_1\cos k-2t_2 \cos 2k$ and interaction strength
$G>0$. The interaction is modulated by the potential
\begin{eqnarray}
\eta_k =\sin (k/2)\sqrt{t_1 + 4 t_2 \cos^2 (k/2)},\label{etakdef}
\end{eqnarray}
which displays the odd-parity behavior $\eta_k=-\eta_{-k}$ characteristic of
$p$-wave superconductivity. The pair potential is related to the single-particle spectrum by
$4\eta_k^2=\varepsilon_k+2 t_+$,  $t_+=t_1+t_2$. This relation is the key to achieve
exact solvability.

In real space, we define $c_j=L^{-1/2}\sum_{k \in {\cal S}_{k}^{\phi}} e^{\im j  k}\, 
\hat{c}_{k}$ for a chain of length $L$, measured in units of the lattice constant. 
We take $\phi$-dependent boundary conditions $c_{i+L}=e^{\im \phi/2}c_i$. 
In a ring geometry, periodic boundary conditions ($\phi=0$) correspond to enclosed 
flux $\Phi=0$ and antiperiodic boundary conditions ($\phi=2\pi$) correspond to $\Phi=\Phi_0$.
The resulting sets of allowed momenta \({\cal S}_{k}^{\phi}\) are   
${\cal S}_{k}^{0}={\cal S}_{k+}^{0}\oplus {\cal S}_{k-}^{0}\oplus \{0,-\pi\}$ 
and 
${\cal S}_{k}^{2\pi}={\cal S}_{k+}^{2\pi}\oplus {\cal S}_{k-}^{2\pi}$,
with 
${\cal S}_{k\pm}^{0}=L^{-1}\{\pm 2\pi, \pm 4\pi,
\cdots, \pm(\pi L-2\pi)\}$ 
and  ${\cal S}_{k\pm}^{2\pi}=L^{-1}\{\pm\pi, \pm 3\pi,
\cdots, \pm(\pi L -\pi)\}$. 
The RGK Hamiltonian in real space reads
\begin{eqnarray}
H_{\sf RGK}&=&-\sum_{i=1}^L \sum_{r=1}^2\left(t_r \, c_{i}^{\dagger}c_{i+r}^{\;}
+{\rm H.c.}\right) - \ 2G \, I^\dagger_\phi I^{\;}_\phi,\\
I_\phi&\equiv& 2 \im \sum_{k \in {\cal S}_{k+}^{\phi}}
\eta_k \ \hat{c}_k\hat{c}_{-k}=\sum_{i>j}^L  \eta(i-j) \, c_{i}^{\;}c_{j}^{\;}.
\end{eqnarray}

There are at least two cases where the pairing function $\eta(m)$ 
can be determined in closed form by Fourier transformation of Eq.\ \eqref{etakdef}. For $t_1=0$ and $t_2\neq 0$,
$\eta(m)=\sqrt{t_2} \, \delta_{m1}$, i.e. we have nearest-neighbor pairing only.
For $t_1\neq 0$ and $t_2= 0$ we obtain
\begin{eqnarray}
\eta(m)=\frac{(-1)^m \, 8 \sqrt{t_1}}{\pi}\frac{m}{1-4m^2} ,\;\;{\rm for}\;\;L\rightarrow\infty,\label{etamdef}
\end{eqnarray}
so a long-range pairing interaction with a slow $1/m$ decay with distance $m=i-j$. In general, $\eta(m)$ is 
a monotonous, decaying function of $m$.

This long-range pairing interaction is the difference with the original Kitaev 
model \cite{Kitaev2001} that allows for the exact solution beyond the mean-field 
approximation. As we shall see in a moment, the long-range coupling still allows 
for a topologically nontrivial phase. It may also be physically relevant for chains 
of magnetic nanoparticles on a superconducting substrate \cite{Cho11,Nad13}, 
which have recently been shown to support topologically protected Majorana zero-modes 
in the presence of a long-range coupling \cite{Pie13}. 


{\it Mean-field approximation.--- }
Before we work out the exact solution of the RGK chain, we would like
to establish first whether it displays a non-trivial topological phase
in the mean-field approximation. 

We set \(t_2=0\) for simplicity and take the pairing interaction \eqref{etamdef}.
The mean-field approximation to the RGK chain is obtained from the
substitution $2GI^\dagger I^{\;}\rightarrow \Delta^* I^{\;} +\Delta \,  
I^\dagger$, with gap function
$\Delta=2G\langle I^{\;}_{L}\rangle=e^{\im \theta} |\Delta|$.
We define Majorana fermion operators
$a_i=e^{-\im \theta/2}c^{\;}_i + e^{\im \theta/2}c^\dagger_i$,
$\im b_i=e^{-\im \theta/2}c^{\;}_i - e^{\im \theta/2}c^\dagger_i$.
The mean-field Hamiltonian is
\begin{equation}
H_{\sf mf}=
\frac{\im t_1}{2}\sum_{i=1}^{L-1}(b_ia_{i+1}-a_ib_{i+1})-\frac{\im}{2} \sum_{i>j}^{L}\Delta_{i-j}(b_ia_{j}+a_ib_{j}),
\end{equation}
where $\Delta_{i-j}=|\Delta| \eta(i-j)$, and displays a topological phase 
characterized by power-law Majorana edge modes and an associated
\(4\pi\)-periodic Josephson effect. For clarity of presentation, 
let us compute approximate edge modes to
leading order in $\delta_2=\Delta_2/(\Delta_1+t_1)$. The Majorana mode localized at the left end of the chain is
$\tilde{a}_1=a_1+\sum_{i=2}^{L-1}\delta_i \sum_{j=2}^{L}(-\delta_2)^{L-j} \, a_j$,
so that
$\im[H_{b}^{\sf mf}, \tilde{a}_1]=\Delta_{L-1}\ b_L+{\cal O}(\delta_2)$.
We see that $\Delta_1+t_1$ controls the localization of the Majoranas, while
\(\Delta_{L-1}\) controls the vanishing of the commutator with the
Hamiltonian.

Due to the long-range pairing interaction \eqref{etamdef}, the wave function of the 
Majorana modes decays algebraically rather than exponentially in the bulk, and their 
energy approaches zero as a power law in $1/L$, similarly to what has been found in 
other mean-field models based on the Kitaev chain with long-range coupling \cite{Pie13,Vod14,DeG13}.

{\it Exact solution.---} To show that the RGK chain is exactly solvable, 
let us rewrite
\begin{align} \!\!\!\!\!\!
H_{\sf RGK}={}& 8 H_{\phi} +\delta_{\phi,0} \, (\varepsilon_0 \hat{c}^\dagger_0 \hat{c}_{0}^{\;}+
\varepsilon_{-\pi} \hat{c}^\dagger_{-\pi}\hat{c}_{-\pi}^{\;}) \nonumber \\
&-4 t_+  \, S^z+C_\phi ,\\
H_\phi={}&\sum_{k \in {\cal S}_{k+}^{\phi}}\eta_{k}^2 \ S_{k}^{z}-G \!\!\!\!\sum_{k,k' \in
{\cal S}_{k+}^{\phi} } \!\!\! \eta_{k}\eta_{k^{\prime}
} \ S_{k}^{+}S_{k^{\prime}}^{-},
\label{HamiltRG0}
\end{align}
with $C_\phi= 2 t_2 \delta_{\phi,0}$. We introduced the operators
$S_{k}^{z}=\frac{1}{2}(  \hat{c}^\dagger_k \hat{c}_{k}^{\;}
+\hat{c}^\dagger_{-k}\hat{c}_{-k}^{\;}-1)$ and
$S_{k}^{+}=\hat{c}_{k}^\dagger \hat{c}_{-k}^\dagger$
for each pair \((k,-k)\) of pairing-active momenta. They satisfy the algebra of
\({\rm SU}(2)\). Hence $S^z= \sum_{k \in {\cal S}_{k+}^{\phi}} S_{k}^{z}$
defines a conserved quantity, \([H_\phi,S^z]=0\).

The Hamiltonian $H_\phi$ belongs to the hyperbolic family 
of exactly solvable pairing Hamiltonians \cite{Duk04,ortiz2005}, whose 
best known representative is the chiral $p$-wave superfluid 
\cite{Sierra2009, rombouts2010,Links2010,Lerma2011}. The rational family includes 
$s$-wave pairing
and has been used in the study of the BCS-BEC crossover 
phenomenon \cite{Ortiz2005}.  Eigenstates for $2M+N_\nu$ fermions are
\begin{eqnarray}
|\Phi_{M,\nu}\rangle =\prod_{\alpha=1}^M \biggl( \sum_{k \in {\cal S}_{k+}^{\phi}}
\frac{\eta_k}{\eta_k^2-E_\alpha} \hat{c}_{k}^\dagger \hat{c}_{-k}^\dagger \biggr)
| \nu\rangle .
\label{wavefunction}
\end{eqnarray}
The state $|\nu\rangle$ with $N_\nu$ unpaired fermions satisfies
$S^-_k| \nu\rangle=0$ for all $k$. Moreover, 
$S^z_k| \nu\rangle=-s_k | \nu\rangle$, with $s_k=0$ if the level 
$k$ is singly-occupied or $s_k=1/2$ if it is empty. The corresponding energy levels are
$E_{M,\nu}=\langle \nu | H_\phi|\nu \rangle+\sum_{\alpha=1}^M E_\alpha$, with
spectral parameters $E_\alpha$ determined by the Richardson-Gaudin equations
\begin{eqnarray}
\sum_{k \in {\cal S}_{k+}^{\phi}}
\frac{s_k}{\eta_k^2-E_{\alpha}}-\sum_{\beta\left(
\neq\alpha\right)  }\frac{1}{E_{\beta}-E_{\alpha}}=\frac{Q_\phi}{E_{\alpha}} ,
\label{Betheeq}
\end{eqnarray}
where $Q_\phi=1/2G-\sum_{k \in {\cal S}_{k+}^{\phi}}s_k+M-1$.

In the case of periodic boundary conditions ($\phi=0$) the two momenta $k=0,-\pi$
are not affected by the interactions and need to be included separately.
The eigenvectors then are $| \Psi_N\rangle=|n_0n_{-\pi}\rangle \otimes |\Phi_{M, \nu}\rangle$,
where  $n_{0},n_{-\pi}\in\{0,1\}$ and the total number of fermions is $N=2M+N_\nu+n_0+n_{-\pi}$.

The Richardson-Gaudin equations \eqref{Betheeq} become singular when two or more 
$E_{\alpha}$'s approach the same single-particle energy $\eta^2_k$ and also around
$E_\alpha=0$. At specific values of the  interaction strength $G$
\begin{equation}
G^\phi_n =\frac{2}{L}g^\phi_n =\frac{2}{L-2(2M+N_\nu)+2(n+1-\delta_{\phi ,0})} ,
\label{couplingn}
\end{equation}
there are $n$ solutions $E_{\alpha}$,  $1 \leq n \leq M$, that vanish 
identically \cite{rombouts2010,Lerma2011}. 
In particular, at $n=M$,
where \eqref{wavefunction} becomes a pair condensate, \(G^\phi_{M}\)
is precisely the Moore-Read coupling. And, at $n=1$ where
$Q_\phi=0$, \(G^\phi_{1}=G_c\) is 
the Read-Green coupling, associated to the non-analytic behavior 
of observables in the thermodynamic limit \cite{rombouts2010,Lerma2011}.

The numerical solution of Eq.\ \eqref{Betheeq}
is particularly simple for values of $G>G^\phi_1$, when all $E_\alpha$'s are
real and negative, and for $G=G^\phi_n$ when $n$ of the $E_\alpha$'s
vanish while $M-n$ are real and negative. We carried
out computations with systems of up to $L\approx 2000$ sites at quarter filling. 
Without the integrability condition, this would have required diagonalization of 
a Hamiltonian matrix with the unwieldy dimension $5\cdot 10^{242}$.
The exact solvability reduces the complexity of the problem to the solution of the
$M \simeq 250$ nonlinear coupled equations \eqref{Betheeq}.

We were also able to perform extrapolations to the thermodynamic limit
$N,L\rightarrow \infty$ at finite density $\rho=N/L$ and rescaled interaction
strength $g=G L/2$. In that limit, Eqs. \eqref{Betheeq} relate to the mean-field 
gap and number equations for $H_\phi$ \cite{rombouts2010}
\begin{align}
&\frac{2\pi}{g}=\int_{0}^{\pi}\frac{\eta_{k}^{2}}{E_{k}} \, dk,\;\;
\rho=\frac{1}{\pi}\int_{0}^{\pi}v_{k}^{2} \, dk,\\
&E_k=\sqrt{\left(\frac{\eta^2_k}{2}-\mu\right)^2+\eta^2_k \Delta^2},\;\;
v^2_k=\frac{1}{2}-\frac{\eta^2_k-2\mu}{4E_k},
\end{align}
with quasi-energies $E_k$ and occupation probabilities $v_k^2$.

{\it Phase diagram and topological transition.---}
To establish the quantum phase diagram of the RGK chain, one needs the ground state 
energy ${\cal E}_0(\rho,g)$ of $H_{\sf RGK}$. Depending on the boundary condition and 
fermion-number parity, one has to consider either $N_\nu=0$ or 1. For periodic boundary conditions, since the
levels $k=0,-\pi$ decouple from the rest, $N_\nu=0$ for both even and odd $N$. If $N$ is odd, 
 the unpaired particle occupies the $k=0$ level without blocking an active
level. For antiperiodic boundary conditions the ground state
has $N_\nu=0$ for $N$ even, while for $N$ odd it has $N_\nu=1$ 
with blocked level $k_0$. The resulting ground state energy is given by
\begin{equation}
{\cal E}^{\phi}_0(N)= 8\sum_{\alpha=1}^M \! E_\alpha- 4 t_+  
\,M + J_{\phi,0} + \delta_{N_\nu,1} (4 \eta_{k_0}^2-2t_+),
\end{equation}
where $J_{\phi,0}=\delta_{\phi,0} \, (\varepsilon_0  \, \delta_{n_0,1}+
\varepsilon_{-\pi}\, \delta_{n_{-\pi},1})$. In the thermodynamic limit the energy 
density reduces to
\begin{equation}
e_0\equiv\lim_{L\rightarrow\infty}{\cal E}^\phi_0/L=-2t_+
\rho-\frac{4}{g}\Delta^2+\frac{4}{\pi}\int^{\pi}_0 \eta^2_k v^2_k \, dk.
\end{equation}

\begin{figure}[tb]
\centerline{\includegraphics[width=0.99\columnwidth]{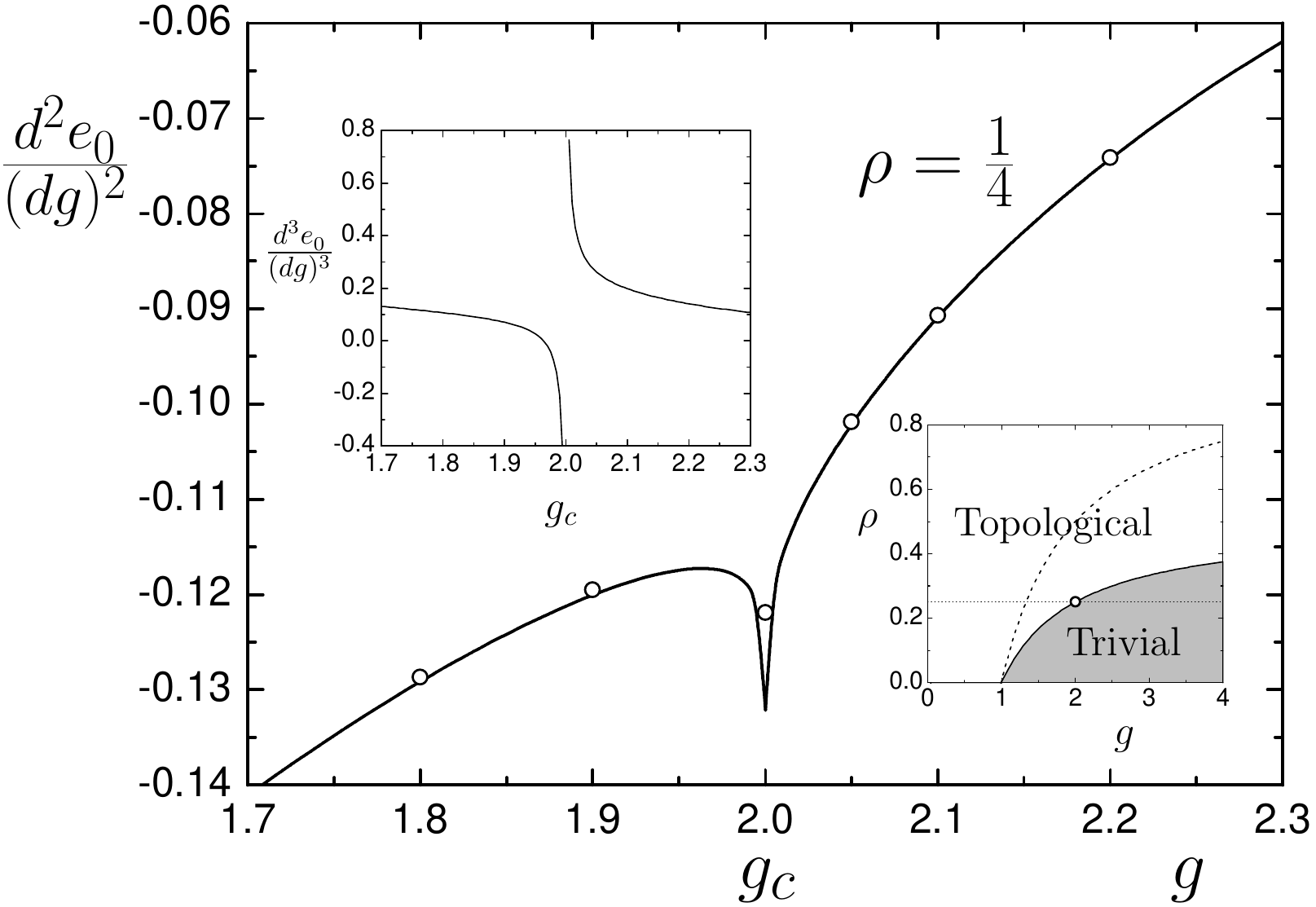}}
\caption{Third-order phase transition between the topological (weak-pairing) and trivial
(strong-pairing) superconducting
phases. The continuous line denotes the second order derivative
of the ground-state energy density $e_0$, evaluated in the thermodynamic limit.
Circles are the exact solution for $L=2048$, $N=512$, and antiperiodic boundary conditions. 
Top inset: Discontinuity in the third-order derivative.
Bottom inset: Quantum phase diagram  in the $(\rho,g)$-plane.
Dashed and full lines represent, respectively, the Moore-Read ($g_M^{-1}=1-\rho$) 
and Read-Green ($g_c^{-1}=1-2\rho$) boundaries.}
\label{fig2}
\end{figure}

The resulting phase diagram is shown in Fig.\ \ref{fig2}.
The RGK chain is gapped for all $g>0$, except for the Read-Green 
coupling $g_c=G^\phi_{n=1} L/2$ where
it becomes critical in the thermodynamic limit (without any 
dependence on the choice of boundary conditions). This critical line defines
the phase boundary separating weak (topological) from strong (trivial) pairing phases,
and thus is a line of non-analyticities. At \(g_c\) a cusp
develops in the second derivative of $e_0$, that leads to
a singular discontinuous behavior of the third-order derivative. Hence the transition
from a weakly-paired to a strongly-paired superconductor is
of third order, just like for the two-dimensional chiral $p$-wave
superconductor \cite{rombouts2010,Lerma2011}. 

Open circles in Fig.\ \ref{fig2} correspond to the second-order derivative
of the exact $e_0$ for the antiperiodic RGK chain with $L=2048$, $N=512$, and 
$t_1=1, t_2=0$, obtained by solving Eqs.\ (\ref{Betheeq}) for some selected pairing strength
values, and illustrates how close to the thermodynamic limit these systems sizes are.
We will demonstrate shortly that the weak-pairing phase of the RGK chain is indeed topologically nontrivial.

{\it Fermion parity switches.---}
We next introduce a quantitative criterion to establish emergence of
topological superconductivity in particle-number conserving,
many-body systems. The criterion exploits the behavior of the
ground state energy of a system of $N$,  and $N\pm1$ particles,
for both periodic and antiperiodic boundary conditions. 
The emergence of topological order in a superconducting wire, closed in a ring 
and described in mean-field, is associated 
 with switches in the ground-state fermion parity ${\cal P}(\phi)$ 
upon increasing the enclosed flux $\Phi=(\phi/2\pi)\times \Phi_0$ 
\cite{Kitaev2001,Kes13,Bee13,Sau13,Hai14,Cre14,Bee14}. 
Any spin-active superconductor, topologically trivial or not, 
may experience a crossing of the ground state energies for even and odd 
number of electrons \cite{Sak70,Bal06,Cha12,Lee14}. Regardless of spin, what matters is the 
number of crossings $N_{X}$ between $\Phi=0=\phi$ and $\Phi=\Phi_0$, $\phi=2\pi$. 
The superconductor is topologically nontrivial if $N_{X}$ is odd, otherwise it is trivial.

In the many-body, number conserving, case we need to identify the relevant parity switches 
signaling the emergence of a topological superconducting phase. Our exact solution 
gives us access to ${\cal P}(\phi)$ only at $\phi=0$ and $\phi=2\pi$, but this is sufficient to determine 
whether $N_X$ is even or odd. Notice that odd $N_X$ means that the flux $\Phi$ should be advanced 
by $2\Phi_0$ --- rather than $\Phi_0$ --- in order to return to the initial ground state, which is the essence of 
the $4\pi$-periodic Josephson effect \cite{Kitaev2001,Sen01}.

\begin{figure}[tb]
\centerline{\includegraphics[width=0.99\columnwidth]{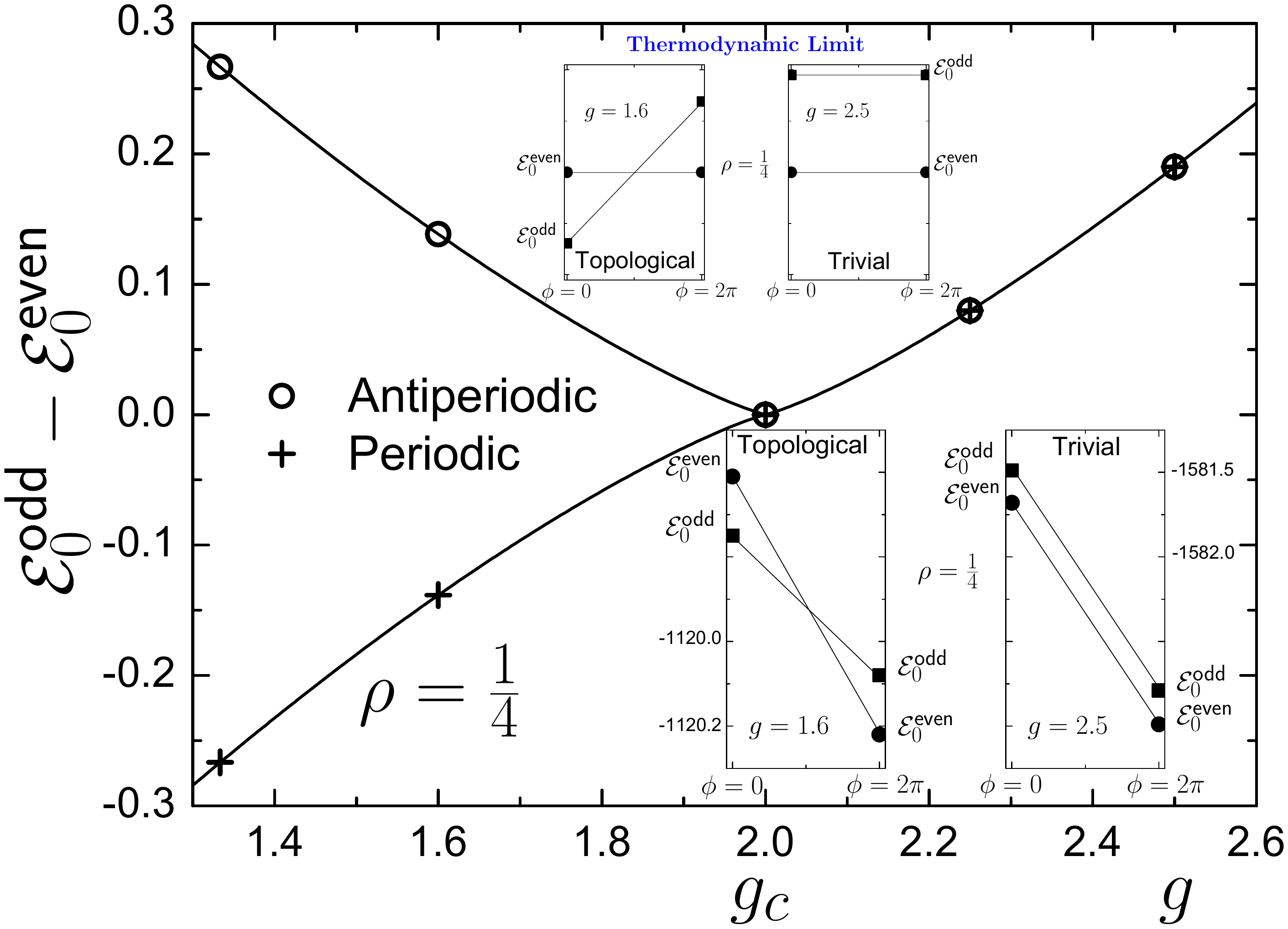}}
 \caption{Ground state energies for the RGK chain (in units of $t_1\equiv 1$, for $t_2=0$) for even 
 ($N=2M)$ and odd ($N=2M\pm 1$) number of fermions, and  with 
 periodic ($\phi=0$) or antiperiodic ($\phi=2\pi$) boundary conditions. The main plot 
 shows the odd-even difference as a function of the interaction strength $g$ for a finite 
 system (data points, for $N=512$, $L=2048$) and in the thermodynamic limit (continuous lines). 
 The topologically nontrivial state is entered for $g<g_c=2$. The insets show the even and odd 
 energies themselves, for the finite system (lower two insets) and in the thermodynamic limit 
 (upper two insets), illustrating the fermion parity switches ${\cal P}_N(\phi)$.}
 \label{fig3}
\end{figure}

To identify the fermion parity switches we calculate the ground state energy 
${\cal E}_0^\phi(N)$ for a given number $N$ of fermions in the chain of length $L$, 
with periodic ($\phi=0$) or antiperiodic ($\phi=2\pi$) boundary conditions, and 
compare ${\cal E}_0^{\sf odd}(\phi)=\tfrac{1}{2}{\cal E}_0^\phi(N+1)+\tfrac{1}{2}{\cal E}_0^\phi(N-1)$ 
and ${\cal E}_0^{\sf even}(\phi)={\cal E}_0^\phi(N)$, where we assumed $N$ even.  
The difference (inverse compressibility) 
$\chi(\phi)= {\cal E}_0^{\sf odd}(\phi)-{\cal E}_0^{\sf even}(\phi)$ determines 
${\cal P}_N(\phi)={\rm sign}\,\chi(\phi)$, so it has the opposite sign at $\phi=0$ and $\phi=2\pi$ 
in the topologically nontrivial phase. We also find that ${\cal P}_{N\in {\sf even}}(\phi)=-{\cal P}_{N\in {\sf odd}}(\phi)$
in the topologically nontrivial phase. 
The results, shown in Fig.\ \ref{fig3}, unambiguously demonstrate the topologically nontrivial 
nature of the superconductor for $g<g_c$ --- both in a finite system and in the thermodynamic limit, 
and without relying on any mean-field approximation. 

The ground state of the odd ($2M\pm1$) system strongly depends on the boundary 
conditions. For periodic boundary conditions the unpaired particle always occupies the 
$k_0=0$ level, while for the antiperiodic case it starts blocking the Fermi 
momentum $k_F=k_0$ at $g=0$, continuously decreasing its modulus with increasing $g$, 
up to $k_0=\pi/L$ at $g_0\sim 1.1936$ ($\rho=1/4$), corresponding to $\mu = \Delta^2$ 
in the thermodynamic limit. 
In that limit $\chi(\phi)$ has a  particularly simple form: 
$\chi(0)=-8\mu$, and $\chi(2\pi)=8 |\mu|$ for $g>g_0$. 

{\it Topological invariant and zero-modes---}
The occupation number \({\cal N}_{k=0}\) of the $k=0$ single-particle state
is a topological invariant \cite{rombouts2010,botelho2005}, being the one-dimensional analogue 
of the winding number in two space dimensions \cite{Volovik1988,Foster2013}. By combining the 
integrals of motion \cite{ortiz2005,rombouts2010} with the Hellmann-Feynman theorem we find 
\begin{align}
&{\cal N}_k=\tfrac{1}{2}-s_k  -4s_k\gamma^2  \sum_{\alpha=1}^M
\frac{\eta_{k}^2}{(\eta_{k}^2-E_\alpha)^2}\frac{\partial E_\alpha}{\partial\gamma} ,\\
&\gamma=\frac{G_c \, W(G)}{3 W(G) - 2 G_c}, \;\;
W(G)=\frac{G_c \,G}{2G- G_c}.
\end{align}
The mapping $W(G)$ is an involution, $W(W(G))\!=\!G$, with fixed point $W(G_c)\!=\!G_c$, 
typical of self-dual transformations. 
In the thermodynamic limit, $\lim_{L\rightarrow\infty}{\cal N}_{k=0}=\Theta[G_c - G]$ with 
$\Theta[x]$ the unit step function, thus signaling the topological phase transition at $G\!=\!G_c$. 

We saw in the mean-field calculation that the Majorana zero-modes have a power-law decay, 
so for any finite $L$ they overlap and are displaced from precisely zero energy. It is known for the 
mean-field Kitaev chain that it is still possible to construct exact zero-energy modes by tuning 
the flux to a fermion parity switch \cite{Akh11}. We may follow a similar strategy for our many-body system. 
By varying the boundary conditions in the topologically nontrivial phase the ground state energies 
\({\cal E}_0^{\sf odd}(\phi)\) and \({\cal E}_0^{\sf even}(\phi)\) cross at some critical $\phi^{\ast}$. 
(They must cross, because their order is inverted at $\phi=0$ and $\phi=2\pi$.) At this value of $\phi=\phi^*$ 
one can identify many-body Majorana zero-modes as follows.

We define the normalized ground states  
$|\Psi_0^{\sf even} \rangle=|\Psi_0^{N} \rangle$,
$|\Psi_0^{\sf odd} \rangle=(|\Psi_0^{N+1} \rangle+|\Psi_0^{N-1} \rangle)/\sqrt{2}$,
with the transition operator
\(\hat{T}=|{\Psi_0^{\sf even}}\rangle \langle \Psi_0^{\sf odd}|\).
Clearly,  $\hat{T}^2=0$, and $\{\hat{T},\hat{T}^\dagger\}=\hat{P}_0$.
Here, $\hat{P}_0=|{\Psi_0^{\sf even}}\rangle \langle \Psi_0^{\sf even}|+
|{\Psi_0^{\sf odd}}\rangle \langle\Psi_0^{\sf odd}|$ is the projector onto the
ground-state subspace. The corresponding Majorana operators are
\begin{eqnarray}
\Gamma_1=\hat{T}+\hat{T}^\dagger \ , \
\im \, \Gamma_2=\hat{T}-\hat{T}^\dagger  \ , \ \{\Gamma_1,\Gamma_2\}=0 .
\end{eqnarray}
We expect these zero-modes to be localized
at the edges of the RGK chain due to the energy gap in the bulk, and it would be interesting 
to check this numerically.

In conclusion, we have constructed a variation of the Kitaev Hamiltonian that is both 
number conserving and interacting, but still exactly solvable. This allowed us to 
identify the fermion parity switches needed for 
characterizing topological superconductivity in interacting many-body systems. 
We have shown that 
our Richardson-Gaudin-Kitaev model shares the features of the mean-field Kitaev 
model that have made it a paradigm of topological superconductivity. There is one difference, 
the long-range nature of the pairing interaction, but in view of recent experimental 
developments \cite{Nad13,Pie13}, this may be a welcome feature of the model rather than a drawback. 
Irrespective of the potential application of our model to a real physical system, we anticipate that its 
integrability will make it a fruitful starting point for theoretical studies of interacting models with topological order.

JD and CE are supported by grant FIS2012-34479 of the Spanish Ministry of Economy
and Competitiveness. EC and CB acknowledge support by the Netherlands Organization 
for Scientific Research (OCW/NWO/FOM) and an ERC Synergy Grant.


\begin{thebibliography}{}

\bibitem{Has10} 
M. Z. Hasan and C. L. Kane, Rev. Mod. Phys. \textbf{82}, 3045 (2010).

\bibitem{Qi11} 
X.-L. Qi and S.-C. Zhang, Rev. Mod. Phys. \textbf{83}, 1057 (2011).

\bibitem{Alicea2012}
J. Alicea, Rep. Prog. Phys. {\bf 75}, 076501 (2012).

\bibitem{Beenakker2013}
C. W. J. Beenakker, Annu. Rev. Con. Mat. Phys. {\bf 4}, 113 (2013).

\bibitem{Ryu2010}
S. Ryu, A. P. Schnyder, A. Furusaki, and A. W. W. Ludwig, New J. Phys. {\bf 12}, 065010 (2010).

\bibitem{Kitaev2001}
A. Yu. Kitaev, Phys. Usp. \textbf{44} (suppl.), 131 (2001).

\bibitem{DeG66} 
P.-G. de Gennes, \textit{Superconductivity of Metals and Alloys} (Benjamin, New York, 1966).

\bibitem{vanHeck2014}
B. van Heck, E. Cobanera, J. Ulrich, and F. Hassler,
Phys. Rev. B {\bf 89}, 165416 (2014).

\bibitem{Mourik2012}
V. Mourik, K. Zuo, S. M. Frolov, S. R. Plissard, E. P.  A. M. Bakkers, and
L. P. Kouwenhoven, Science {\bf 336}, 1003 (2012).

\bibitem{Das2012}
A. Das, Y. Ronen, Y. Most, Y. Oreg, M. Heiblum, and H. Shtrikman,
Nature Phys.\ {\bf 8}, 887 (2012).

\bibitem{Deng2012}
M. T. Deng, C. L. Yu, G. Y. Huang, M. Larsson, P. Caroff, and H. Q. Xu,
Nano Lett.\ {\bf 12}, 6414 (2012).

\bibitem{Fin13} 
A. D. K. Finck, D. J. Van Harlingen, P. K. Mohseni, K. Jung, and X. Li, Phys. Rev. Lett. \textbf{110}, 126406 (2013).

\bibitem{Chu13} 
H. O. H. Churchill, V. Fatemi, K. Grove-Rasmussen, M. T. Deng, P. Caroff, H. Q. Xu, 
and C. M. Marcus, Phys. Rev. B \textbf{87}, 241401 (2013).

\bibitem{Sta13} 
T. D. Stanescu and S. Tewari, J. Phys. Cond. Matt. \textbf{25}, 233201 (2013).

\bibitem{Tsvelik}
A. M. Tsvelik, arXiv:1106.2996.

\bibitem{Cheng2011}
M. Cheng and H.-H. Tu, Phys. Rev. B {\bf 84}, 094503 (2011).

\bibitem{Hassler2012}
F. Hassler and D. Schuricht, New J. Phys. {\bf 14}, 125018 (2012).

\bibitem{Manolescu2014}
A. Manolescu, D. C. Marinescu, and T. D. Stanescu, J. Phys.
Condens. Matter {\bf 26}, 172203 (2014).

\bibitem{Sau2011}
J. D. Sau, B. I. Halperin, K. Flensberg, and S. Das
Sarma, Phys. Rev. B {\bf 84}, 144509 (2011).

\bibitem{Fidkowski2011}
L. Fidkowski, R. M. Lutchyn, C. Nayak, and M. P. A.
Fisher, Phys. Rev. B {\bf 84}, 195436 (2011).

\bibitem{vanHeck2011}
B. van Heck, F. Hassler, A. R. Akhmerov, and C. W. J. Beenakker, 
Phys. Rev. B {\bf 84}, 180502(R) (2011).

\bibitem{Amico2001}
 L. Amico, A. Di Lorenzo, and A. Osterloh, Phys. Rev. Lett. {\bf 86}, 5759  (2001).
 
\bibitem{Dukelsky2001}
J. Dukelsky, C. Esebbag, and P. Schuck, Phys. Rev. Lett. {\bf 87}, 066403 (2001).

\bibitem{ortiz2005}
G. Ortiz, R. Somma, J. Dukelsky, and S. M. A. Rombouts, Nuc. Phys. B {\bf 707}, 421 (2005).

\bibitem{Richardson1963}
R. W. Richardson, Phys. Lett.  {\bf 3}, 277 (1963).

\bibitem{Gaudin1976}
M. Gaudin, J. Phys. (Paris) \textbf{37}, 1087 (1976).

\bibitem{Cho11} 
T.-P. Choy, J. M. Edge, A. R. Akhmerov, and C. W. J. Beenakker, Phys. Rev. B \textbf{84}, 195442 (2011).

\bibitem{Nad13} 
S. Nadj-Perge, I. K. Drozdov, B. A. Bernevig, and A. Yazdani, Phys. Rev. B \textbf{88}, 020407 (2013).

\bibitem{Pie13} 
F. Pientka, L. I. Glazman, and F. von Oppen, Phys. Rev. B \textbf{88}, 155420 (2013).

\bibitem{Vod14} 
D. Vodola, L. Lepori, E. Ercolessi, A. V. Gorshkov, and G. Pupillo, arXiv:1405.5440.

\bibitem{DeG13} 
W. DeGottardi, M. Thakurathi, S. Vishveshwara, and D. Sen, Phys. Rev. B \textbf{88}, 165111 (2013).

\bibitem{Duk04} 
J. Dukelsky, S. Pittel, and G. Sierra, Rev. Mod. Phys. \textbf{76}, 643 (2004).

\bibitem{Sierra2009}
M. I. Iba\~{n}ez, J. Links, G. Sierra, and S. Y. Zhao,
Phys. Rev. B {\bf 79}, 180501(R) (2009).

\bibitem{rombouts2010}
S. M. A. Rombouts, J. Dukelsky, and G. Ortiz, Phys. Rev. B {\bf 82}, 224510 (2010).

\bibitem{Links2010}
C. Dunning, M. Ibanez, J. Links, G. Sierra, and S.-Y. Zhao, J. Stat. Mech. P08025 (2010).

\bibitem{Lerma2011}
S. Lerma H., S. M. A. Rombouts, J. Dukelsky, and G. Ortiz,
Phys. Rev. B {\bf 84}, 100503(R) (2011).

\bibitem{Ortiz2005}
G. Ortiz and J. Dukelsky, Phys. Rev. A {\bf 72}, 043611 (2005). 



\bibitem{Kes13} 
A. Keselman, L. Fu, A. Stern, and E. Berg, Phys. Rev. Lett. \textbf{111}, 116402 (2013).

\bibitem{Bee13} 
C. W. J. Beenakker, J. M. Edge, J. P. Dahlhaus, D. I. Pikulin, S. Mi, and 
M. Wimmer, Phys. Rev. Lett. \textbf{111}, 037001 (2013).

\bibitem{Sau13} 
J. D. Sau and E. Demler, Phys. Rev. B \textbf{88}, 205402 (2013)

\bibitem{Hai14} 
A. Haim, A. Keselman, E. Berg, and Y. Oreg, Phys. Rev. B \textbf{89}, 220504 (2014).

\bibitem{Cre14} 
F. Cr\'{e}pin and B. Trauzettel, Phys. Rev. Lett. \textbf{112}, 077002 (2014).

\bibitem{Bee14} 
C. W. J. Beenakker, arXiv:1407.2131.

\bibitem{Sak70} 
A. Sakurai, Prog. Theor. Phys. \textbf{44}, 1472 (1970).

\bibitem{Bal06} 
A. V. Balatsky, I. Vekhter, and J.-X. Zhu, Rev. Mod. Phys. \textbf{78}, 373 (2006).

\bibitem{Cha12} 
W. Chang, V. E. Manucharyan, T. S. Jespersen, J. Nyg{\aa}rd, and C. M. Marcus, Phys. Rev. Lett. \textbf{110}, 217005 (2013).

\bibitem{Lee14} 
E. J. H. Lee, X. Jiang, M. Houzet, R. Aguado, C. M. Lieber, and S. De Franceschi, Nature Nanotech. \textbf{9}, 79 (2014).

\bibitem{Sen01} 
K. Sengupta, I. Z\v{u}ti\'{c}, H.-J. Kwon, V. M. Yakovenko, and S. Das Sarma, Phys. Rev. B \textbf{63}, 144531 (2001).

\bibitem{botelho2005}
S. S. Botelho and C. A. R. S\`{a} de Melo, J. Low Temp. Phys. {\bf 140}, 409 (2005).

\bibitem{Volovik1988}
G. E. Volovik, Sov. Phys. JETP {\bf 67}, 1804 (1988).

\bibitem{Foster2013}
M. S. Foster, M. Dzero, V. Gurarie, and E. A. Yuzbashyan,
Phys. Rev. B {\bf 88}, 104511 (2013).

\bibitem{Akh11} 
A. R. Akhmerov, J. P. Dahlhaus, F. Hassler, M. Wimmer, and 
C. W. J. Beenakker, Phys. Rev. Lett. \textbf{106}, 057001 (2011). [Supplemental material.]

\end{thebibliography}
\end{document}